\documentclass[10pt,journal,compsoc]{IEEEtran}
\usepackage{graphicx}
\usepackage{graphics}
\usepackage{
lipsum}
\usepackage{tabularx} 
\usepackage{makecell} 
\usepackage{hhline} 
\usepackage{color,soul} 
\usepackage{paralist} 
\usepackage{algorithm}
\usepackage[noend]{algpseudocode}
\usepackage{amsmath} 
\usepackage{array} 
\usepackage{float}
\usepackage{dblfloatfix}    
\usepackage{enumitem}
\usepackage{bbding} 
\usepackage{subcaption} 
\usepackage{mwe} 
\usepackage[export]{adjustbox}

\ifCLASSOPTIONcompsoc
  \usepackage[nocompress]{cite}
\else
  \usepackage{cite}
\fi

\ifCLASSINFOpdf
\else
\fi

\begin{document}
\title{N-BaIoT: Network-based Detection\\of IoT Botnet Attacks\\Using Deep Autoencoders}

\author{Yair~Meidan, Michael~Bohadana,
         Yael~Mathov, Yisroel~Mirsky,\\Dominik~Breitenbacher, Asaf~Shabtai, and~Yuval~Elovici

}

\markboth{IEEE Pervasive Computing,~Vol.~13, No.~9, July-September~2018}%
{Bohadana \MakeLowercase{\textit{et al.}}: Network-Based Detection of IoT Devices Compromised by Botnets}

\IEEEtitleabstractindextext{%
\begin{abstract}

The proliferation of IoT devices which can be more easily compromised than desktop computers has led to an increase in the occurrence of IoT-based botnet attacks. 
In order to mitigate this new threat there is a need to develop new methods for detecting attacks launched from compromised IoT devices and differentiate between hour and millisecond long IoT-based attacks. 
In this paper we propose and empirically evaluate a novel network-based anomaly detection method which extracts behavior snapshots of the network and uses deep autoencoders to detect anomalous network traffic emanating from compromised IoT devices. 
To evaluate our method, we infected nine commercial IoT devices in our lab with two of the most widely known IoT-based botnets, Mirai and BASHLITE. 
Our evaluation results demonstrated our proposed method's ability to accurately and instantly detect the attacks as they were being launched from the compromised IoT devices which were part of a botnet.
\end{abstract}

}

\maketitle

\IEEEdisplaynontitleabstractindextext
\IEEEpeerreviewmaketitle

\ifCLASSOPTIONcompsoc
\IEEEraisesectionheading{\section{Introduction}
\label{sec:introduction}}
\else
\section{Introduction}
\label{sec:introduction}
\fi
\IEEEPARstart{A}{s} the number of \emph{Internet of Things (IoT)} devices deployed dramatically increases worldwide~\cite{Kolias2017DDoSBotnets
}, and the traffic volume of IoT-based DDoS attacks reaches unprecedented levels~\cite{Kolias2017DDoSBotnets, Bertino2017BotnetsSecurity, Hallman2017IoDDoSBotnets}, the need for timely detection of IoT botnet attacks 
has become imperative for mitigating the risks associated with these attacks. 
Instantaneous detection promotes network security, as it expedites the alerting and disconnection of compromised IoT devices from the network, thus stopping the botnet from propagating and preventing further outbound attack traffic.


\begin{table*}[!htbp]
  \caption{Prior studies conducted on the detection of IoT-related anomalies, botnets, and malware attacks}
  \centering
  \resizebox{\textwidth}{!}{
  \begin{tabular}{c||cccccccc}
    \hhline{=========}
    \textbf{Paper}
    &\begin{tabular}[t]{@{}c@{}}\textbf{Detected}\\\textbf{Botnet}\end{tabular}
    &\begin{tabular}[t]{@{}c@{}}\textbf{Botnet}\\\textbf{Operational}\\\textbf{Step}\end{tabular}
    &\textbf{Attack(s)}
    &\begin{tabular}[t]{@{}c@{}}\textbf{Detection}\\\textbf{Approach}\end{tabular}
    &\begin{tabular}[t]{@{}c@{}}\textbf{Deployment}\\\textbf{Level}\end{tabular}
    &\begin{tabular}[t]{@{}c@{}}\textbf{Assumed}\\\textbf{Environment}\end{tabular}
    &\begin{tabular}[t]{@{}c@{}}\textbf{Research}\\\textbf{Type}\end{tabular}
    &\begin{tabular}[t]{@{}c@{}}\textbf{Data}\\\textbf{for}\\\textbf{Evaluation}\end{tabular}
    \\
    \hhline{=========}
 	\cite{Bertino2017BotnetsSecurity}&\makecell{Linux.Darlloz\\worm, Mirai}&Infection&DDoS&\makecell{Intrusion prevention,\\traffic monitoring}&\makecell{Network\\(routers, gateways)}&-&\makecell{Survey}&-\\
    \hline
    \cite{Hallman2017IoDDoSBotnets}&Mirai&\makecell{Various operational\\steps, depending\\on the malware}&DDoS&-&-&-&Survey&-\\
    \hline
    \cite{Ozcelik2017Software-DefinedDDoS}&Mirai&\makecell{Scanning\\(propagation)}&\makecell{Mirai-infected\\IoT devices scan \\for further devices}&\makecell{Dynamic\\updating\\of flow rules}&"Thin fog"&\makecell{Critical\\ infrastructures}&Experimental&\makecell{Emulated\\IoT nodes,\\simulated data}\\
    \hline
    \cite{Summerville2016Ultra-lightweightDevices}&-&-&\makecell{Worm propagation,\\code injection,\\ tunneling attack}&\makecell{Deep packet\\anomaly detection}&Host&-&Experimental&\makecell{Two real\\devices}\\   
    \hline  
    \cite{Pa2016IoTPOT:Threats}
 &\makecell{ZORRO, *.sh,\\GAFGYT,\\ KOS, nttpd}&All&-&\makecell{Honeypot to\\collect and\\analyze attacks}&Both&-&Experimental&\makecell{Real\\data}\\ 
    \hline       		
    \cite{Sedjelmaci2016AMethodology}&-&-&\makecell{Devices are\\attacked by\\a DoS attack}&\makecell{Hybrid: signature-\\based and anomaly\\detection (BPN)}&Host&WSN&Experimental&Simulation\\
    \hline    
    \cite{Bostani2017HybridApproach}&-&-&\makecell{Routing attacks\\(sinkhole and\\selective-forwarding)}&\makecell{Hybrid: specification-\\based and anomaly\\detection (OFPC)}&\makecell{Network\\(routers and\\root nodes)}&\makecell{6LoWPAN WSN,\\representing\\a smart city}&Experimental&Simulation\\
    \hline
    \cite{Butun2015AnomalyThings}&-&-&-&\makecell{Several methods,\\including\\anomaly detection}&\makecell{Network\\(cloud)}&\makecell{Sensing systems\\and distributed\\cloud platforms}&\makecell{Survey on\\challenges and\\detection approaches}&-\\
    \hline
    \cite{Midi2017KalisThings}&-&-&\makecell{ICMP flood, replication, wormhole,\\TCP SYN flood, HELLO jamming, data\\modification, selective forwarding, smurf}&\makecell{Knowledge\\driven,\\ anomaly detection}&Network&\makecell{Adapts to ZigBee/XBee/\\6LoWPAN (on IEEE 802.15.4),\\WiFi (on IEEE 802.11), and BT}&Experimental&\makecell{Real devices,\\simulated data}\\
    \hline
    \cite{Raza2013SVELTE:Things}&-&-&\makecell{Routing attacks like spoofed\\or altered information,\\sinkhole, selective-forwarding}&\makecell{Hybrid: signature-\\based and\\anomaly detection}&\makecell{Hybrid:\\border router\\and hosts}&6LoWPAN&Experimental&Simulation\\
    \hline
    \cite{2017AThings}&-&-&-&\makecell{Several methods,\\including\\anomaly detection}&\makecell{Host and\\network}&-&Survey&-\\
    \hhline{=========}
  \end{tabular}
  }
  \label{tab:related_work}
\end{table*}

Botnets such as Mirai are typically constructed in several distinct operational steps~\cite{Kolias2017DDoSBotnets}, namely \emph{propagation}, \emph{infection}, \emph{C\&C communication}, and \emph{execution of attacks}. Unlike most previous studies on botnet detection (see Table~\ref{tab:related_work}), which addressed the early operational steps, we focus on the last step. 
We concentrate on large enterprises, which are 
expected to face an ever growing range and quantity of IoT devices, normally connecting to their networks via Wi-Fi (short-range communications like Bluetooth and ZigBee are not in our current scope). These devices can be either self-deployed (e.g., \textit{smart} smoke detectors) or dynamically introduced from the outside by employees and visitors (e.g., BYO wearables). 


Assuming that botnet attacks are unlikely to disappear, the fundamental question we address is as follows. Given a large number of heterogeneous IoT devices connected to an organizational network, can we devise a centralized, automated method that is highly effective and accurate in detecting compromised IoT devices which have been added to a botnet and have been used to launch attacks? 


For detecting attacks launched from IoT bots we propose a network-based approach, which uses deep learning techniques to perform anomaly detection. Specifically, we extract statistical features which capture behavioral snapshots of benign IoT traffic, and train a deep autoencoder \emph{(one for each device)} to learn the IoT's normal behaviors. The deep autoencoder attempts to compress 
snapshots. When an autoencoder fails to reconstruct a snapshot, then it is a strong indication that the observed behavior is anomalous (i.e., the IoT device has been compromised and is exhibiting an unknown behavior). An advantage of using deep autoencoders, is their ability to learn complex patterns, e.g., of various device functionalities
. This results in an anomaly detector with hardly any false alarms. We empirically show that the autoencoders' false alarm rate is considerably lower than three other algorithms commonly used for anomaly detection~\cite{tuor2017deep}.

The following are the benefits of using this approach to detecting infected IoTs:

\textbf{Heterogeneity tolerance}. Compared to classical computing environments, the IoT domain is highly diverse~\cite{Bertino2017BotnetsSecurity, Hallman2017IoDDoSBotnets}. However, by profiling each device with a separate autoencoder, our method addresses the growing heterogeneity of IoT devices.

\textbf{Open World}. Typically in deep learning applications, models are trained to classify based on labels provided by experts (e.g. malicious or benign). However, our autoencoders are trained to detect when a behavior is abnormal. Thus our method can detect new previously 'unseen' botnet behaviors, which is important given the continuously evolving variants~\cite{Bertino2017BotnetsSecurity} or new botnets, which already make most detection methods obsolete~\cite{garcia2014survey}.

\textbf{Efficiency}. In the enterprise scenario, it is common that the traffic data of all connected hosts is monitored, but the amount of monitored traffic is prohibitively large to store and use for training deep neural networks.  Our method uses incremental statistics to perform the feature extraction, and the training of the autoencoders can be performed in semi-online manner (train on a batch of observations and then discard). Therefore the training is practical, and there is no storage concern. Additionally, 
our method is network-based so it does not consume any computation, memory, or energy resources from the (typically constrained) IoT devices. Thus, our method does not jeopardize their functionality or impair their lifespan. Our focus on the attack operational step (as opposed to the early steps) also makes our method indifferent to the botnet propagation protocols 
and the possibly encrypted~\cite{garcia2014survey} C\&C channels.

The contributions of this paper can be summarized as follows:
\begin{enumerate}
	\item To the best of our knowledge, we are the first to apply autoencoders to IoT network traffic for anomaly detection, as a complete means of detecting botnet attacks. Even in the larger domain of network traffic analysis, autoencoders have not been used as fully automated standalone malware detectors, but rather as preliminary tools for either feature learning~\cite{arnaldo2017learning} or dimensionality reduction~\cite{li2015hybrid}, or at most as semi-manual outlier detectors which substantially depend on human labeling for subsequent classification~\cite{veeramachaneni2016ai} or further inspection by security analysts~\cite{tuor2017deep}.
    \item Unlike previous experimental studies on the detection of IoT botnets or IoT traffic anomalies which relied on emulated or simulated data (\hspace{1sp}\cite{Sedjelmaci2016AMethodology, Ozcelik2017Software-DefinedDDoS, Bostani2017HybridApproach, Midi2017KalisThings}), we perform empirical evaluation with real traffic data, gathered from nine commercial IoT devices infected by authentic botnets from two families. We examine Mirai and BASHLITE, two of the most common IoT-based botnets, which have already demonstrated~\cite{Kolias2017DDoSBotnets} their harmful capabilities
. To enable reproducibility and address the lack of public botnet datasets~\cite{garcia2014survey}, particularly for the IoT, we share our network traces at http://archive.ics.uci.edu/ml/datasets/detection\_of\_
IoT\_botnet\_attacks\_N\_BaIoT.
\end{enumerate}

\section{Related Work}\label{sec:related_work}
The botnet detection methods suggested thus far can be categorized based on (1) the specific operational step to be detected, and (2) the detection approach. Table~\ref{tab:related_work} is based on this categorization and further summarizes previous studies on the detection of IoT-related anomalies, botnets, and malware attacks. 

Among the \emph{botnets' operational steps}, previous IoT-related detection studies (e.g., ~\cite{Ozcelik2017Software-DefinedDDoS} and~\cite{Summerville2016Ultra-lightweightDevices}) focused mainly on the early steps of propagation and communication with the C\&C server. 
However, given that botnet attacks continue to mutate on a daily basis~\cite{Kolias2017DDoSBotnets} and become increasingly sophisticated~\cite{Bertino2017BotnetsSecurity}, we anticipate that some of these mutations will eventually succeed at bypassing existing methods of early detection. 
Moreover, mobile IoT devices 
might get contaminated when connected to external networks.
For instance, smartwatches may connect to dubious \emph{free Wi-Fi} networks when their owners arrive at airports. 
Hence, monitoring organizational networks for identifying the early steps of infection alone is insufficient.
Accordingly, we focus on a later step of a botnet operation, when IoT bots begin launching cyberattacks. In that sense, our method adds a \emph{last line of defense} security layer. 
It instantly detects the IoT-based attacks and minimizes their impact by issuing an immediate alert which recommends the isolation of any compromised device from the network until it is sanitized.

\begin{table*}[!t]
  \caption{Extracted features}
  \centering
  \begin{tabular}{lllc}
    \hhline{====}	
     \multicolumn{1}{c}{\textbf{Value}}                                                 & \multicolumn{1}{c}{\textbf{Statistic}}                                              & \multicolumn{1}{c}{\textbf{Aggregated by}}                     & \begin{tabular}[t]{@{}c@{}}\multicolumn{1}{c}{\textbf{Total Number}}\\\multicolumn{1}{c}{\textbf{of Features}}\end{tabular} \\
    \hhline{====}	
    Packet size (of outbound packets only)                    & Mean, Variance                                         & \makecell[l]{Source IP,\textsuperscript{1} Source MAC-IP,\textsuperscript{2}\\Channel, Socket\textsuperscript{3}} & 8                    \\
    \hline
    Packet count                                           & Number                                                 & \makecell[l]{Source IP, Source MAC-IP,\\Channel, Socket} & 4                     \\
    \hline
    \makecell[l]{Packet jitter (the amount of time\\between packet arrivals)} & Mean, Variance, Number                                         & Channel                           & 3                      \\
    \hline
    \makecell[l]{Packet size (of both inbound and\\outbound together)}    & \makecell[l]{Magnitude, Radius, Covariance,\\Correlation coefficient} & Channel, Socket                   & 8                    \\
    \hhline{====}	
    \multicolumn{4}{l}{\textsuperscript{1} The source IP is used to track the host as a whole.}\\
    \multicolumn{4}{l}{\textsuperscript{2} The source MAC-IP adds the capability to distinguish between traffic originating from different gateways and spoofed IP addresses.}\\
    \multicolumn{4}{l}{\makecell[l]{\textsuperscript{3} The sockets are determined by the source and destination TCP or UDP port numbers. For example, all of the traffic sent from\\192.168.1.12:1234 to 192.168.1.50:80 (traffic flowing from one socket to another).}}\\
    \\
    \multicolumn{4}{l}{\makecell[l]{Further details and the datasets themselves are publicly available at\\ http://archive.ics.uci.edu/ml/datasets/detection\_of\_IoT\_botnet\_attacks\_N\_BaIoT.}}\\
    \hhline{====}
    \end{tabular}
  \label{tab:extracted_features}
\end{table*}

Among the suggested \emph{botnet detection approaches}, a primary distinction is made between host-based~\cite{Sedjelmaci2016AMethodology, Summerville2016Ultra-lightweightDevices} and network-based~\cite{Ozcelik2017Software-DefinedDDoS, Bostani2017HybridApproach, Butun2015AnomalyThings, Midi2017KalisThings} approaches. We consider host-based techniques less realistic for detecting compromised IoT devices, because (1) we cannot rely on the good will of all IoT manufacturers to install designated host-based anomaly detectors on their products; (2) there is limited access to some IoT devices (e.g., wearables), so the installation of software on end devices cannot be enforced; (3) the constrained computation and power of most IoT devices impose constraints on the complexity and efficiency of host-based anomaly detection algorithms, which also might consume energy and computation from the devices and thus harm their functionality; and (4) in the enterprise scenario we assume, where various and numerous IoT devices connect to the organizational network, a single non-distributed solution is preferred.

A hierarchical taxonomy of network-based botnet detection approaches, not limited to the IoT domain, is proposed by~\cite{garcia2014survey}. Honeypots are one of the detection sources surveyed in this study. Honeypots have commonly been used for collecting, understanding, characterizing, and tracking 
botnets~\cite{Pa2016IoTPOT:Threats}
. However, they are not necessarily useful for detecting compromised endpoints or the attacks emanating from them. Moreover, honeypots normally require a substantial investment in procurement or emulation of real devices, 
data inspection, signature extraction, 
and keeping up with mutations. 
As per~\cite{garcia2014survey}, normal networks constitute an alternative detection source, where network intrusion detection systems (NIDSs) monitor traffic data continuously and automatically, while using pattern matching to detect signs of undesirable activities. 
Such patterns may rely on (1) signatures identified by honeypots, (2) DNS traffic with a potential C\&C server, (3) traffic anomalies~\cite{Summerville2016Ultra-lightweightDevices}, (4) data mining, or (5) hybrid approaches~\cite{Sedjelmaci2016AMethodology, Bostani2017HybridApproach}. 
Similar to~\cite{Summerville2016Ultra-lightweightDevices}, we find that the anomaly-based approach is best suited for detecting compromised IoT devices, because these connected appliances are typically task-oriented (e.g., specifically designed to detect motion or measure humidity). 
Accordingly, they execute fewer, and potentially less, complex network protocols, and exhibit traffic with less variance than PCs. As such, detecting deviations from their normal patterns should be more accurate and robust.

Many detection algorithms were surveyed in~\cite{garcia2014survey}, however 
artificial neural networks were left uncited, and autoencoders were not mentioned at all. 
Such works within the greater domain of cybersecurity have been published more recently, yet they are dissimilar to our approach, unrelated to the IoT, and often not directly connected to botnets. For instance,~\cite{arnaldo2017learning, li2015hybrid} and~\cite{yu2017network} applied shallow autoencoders for preliminary feature learning and dimensionality reduction, followed by Random Forest, Deep Belief Networks, and Softmax, respectively for classification and fine-tuning. 
Although autoencoders were extended for outlier detection in~\cite{veeramachaneni2016ai}, they still required security analysts to actively label data for subsequent supervised learning. 
Closer to our approach, the authors of~\cite{tuor2017deep} apply deep learning to system logs for detecting insider threats. 
Differently from us, they use DNNs and RNNs (LSTMs), 
and depend on further manual inspection.

In conclusion, our method differs from previous studies as we learn from benign data by training deep autoencoders for each device, and use them as standalone automatic tools for instantaneous detection of existing and unseen IoT botnet attacks.


\section{Proposed Detection Method}\label{sec:proposed_detection_method}
The method we propose for detecting IoT botnet attacks relies on deep autoencoders for each device, trained on statistical features extracted from benign traffic data. 
When applied to new (possibly infected) data of an IoT device, detected anomalies may indicate that the device is compromised. This method consists of the following main stages: (1) data collection, (2)	feature extraction, (3) training an anomaly detector, 
and (4) continuous monitoring. 

\textbf{Data collection.} We capture the raw network traffic data (in \emph{pcap} format) 
using port mirroring on the switch through which the organizational traffic typically flows. 
To ensure that the training data is clean of malicious behaviors, the normal traffic of an IoT is collected immediately following its installation in the network.

\textbf{Feature extraction.} Whenever a packet arrives, we take a behavioral snapshot of the hosts and protocols that communicated this packet. The snapshot obtains the packet's context by extracting 115 traffic statistics over several temporal windows to summarize all of the traffic that has (1) originated from the same IP in general, (2) originated from both the same source MAC and the same IP address, (3) been sent between the source and destination IPs (\emph{channel}), and (4) been sent between the source to destination TCP/UDP sockets (\emph{socket}).


We extract the same set of 23 features (capturing the above, see Table~\ref{tab:extracted_features}) from five time windows of the most recent 100ms, 500ms, 1.5sec, 10sec, and 1min. 
These features can be computed very fast and incrementally and thus facilitate real time detection of malicious packets. Additionally, although generic these features can capture specific behaviors like source IP spoofing~\cite{Bertino2017BotnetsSecurity}, characteristic of Mirai's attacks. For instance, when a compromised IoT device spoofs an IP, the features aggregated by the Source MAC-IP, Source IP and Channel will immediately indicate a large anomaly due to the unseen behavior originating from the spoofed IP address.

\textbf{Training an anomaly detector.} As our base anomaly detector, we use deep autoencoders and maintain a model for each IoT device separately. An autoencoder is a neural network which is trained to reconstruct its inputs after some compression. The compression ensures that the network learns the meaningful concepts and the relation among its input features. If an autoencoder is trained on benign instances only, then it will succeed at reconstructing normal observations, but fail at reconstructing abnormal observations (unknown concepts). When a significant reconstruction error is detected, then we classify the given observations as being an anomaly.



We optimize the parameters and hyperparameters of each trained model such that when applied to unseen traffic the model maximizes the true positive rate (TPR, detecting attacks once they occur) and minimizes the false positive rate (FPR, wrongly marking benign data as malicious). 
For training and optimization, we use two separate datasets which only contain benign data, from which the model \emph{learns} patterns of normal activity. 
The first dataset is the \emph{training set} (
\(DS_{trn}\)), and it is used for training the 
autoencoder, given input parameters such as the \emph{learning rate} (
\(\eta\), the size of the gradient descent step), and the number of \emph{epochs} (complete passes through the entire \(DS_{trn}\)). 
The second dataset is the \emph{optimization set} (
\(DS_{opt}\)), and it is used to optimize these two hyperparameters (\(\eta\) and \(epochs\)) iteratively until the mean square error (\(MSE\)) between a \(model\)'s input (the original feature vector) and output (the reconstructed feature vector) stops decreasing. Stopping at this point 
prevents overfitting $DS_{trn}$, thus promoting better detection results with future data. \(DS_{opt}\) is later used to optimize a threshold (
$tr$) which discriminates between benign and malicious observations; finally, it is also used to optimize the window size (
$ws$), by which the FPR is minimized.

Once the \(model\) training and optimization is complete the 
\(tr^*\) is set. This anomaly threshold, above which an instance is considered anomalous, is calculated as the sum of the sample mean and standard deviation of \(MSE\) over \(DS_{opt}\) (see Equation~\ref{eq:threshold}).

\begin{equation}
tr^*=\overline{MSE}_{DS_{opt}}+s(MSE_{DS_{opt}})
\label{eq:threshold}
\end{equation}

Preliminary experiments revealed that deciding whether a device's packet stream is anomalous or not based on a single instance enables very accurate detection of IoT-based botnet attacks (high TPR). However, benign instances were too often (in approximately 5-7\% of cases) falsely marked as anomalous. 
Thus we base the abnormality decision on a \emph{sequence} of instances by implementing a majority vote on a moving window. We determine the minimal window size \(ws^*\) as the shortest sequence of instances, a majority vote which produces 0\% FPR on \(DS_{opt}\) (see Equation~\ref{eq:window_size}).
    
 \begin{equation}
 ws^*=\operatorname*{arg\,min}_{|ws|}(|\{packet \in ws|MSE(packet)>tr^*\}|>\frac{|ws|}{2})
 \label{eq:window_size}
 \end{equation}
 
\begin{figure*}[!t]
\centering
\includegraphics[width=0.9\textwidth]
{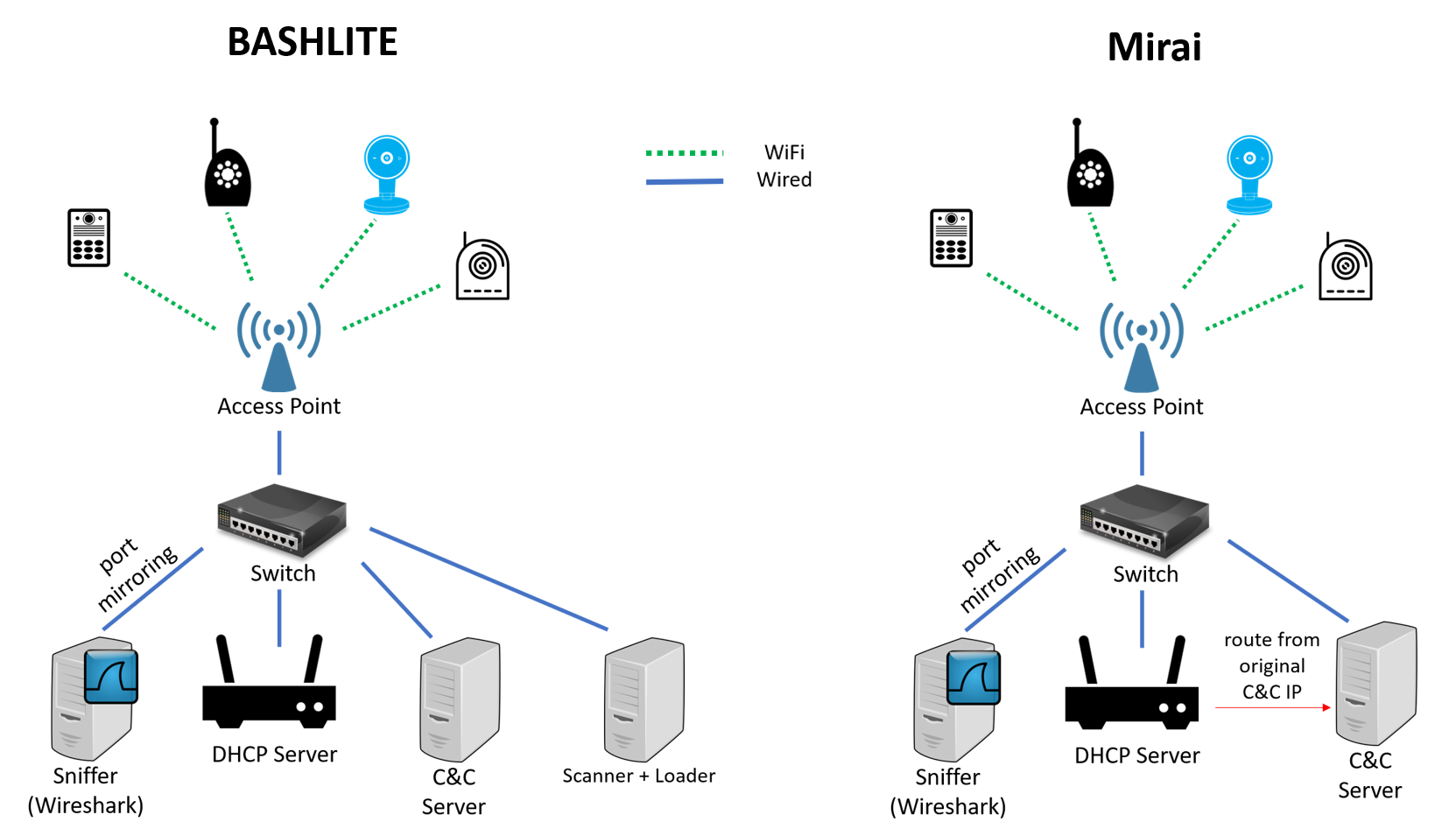}
\caption{Lab setup for detecting IoT botnet attacks} 
\label{fig:lab_architecture}
\end{figure*}

\textbf{Continuous monitoring for anomaly detection.} Eventually, we apply the optimized \(model\) to feature vectors extracted from continuously observed packets
, to mark each instance as benign or anomalous. 
Then, a majority vote on a sequence (the length of \(ws^*\)) of marked instances is used to decide whether the entire respective stream is benign or anomalous. 
Consequently, an alert can be issued upon the detection of an anomalous stream, as it might indicate malicious activity on the IoT device.


\section{Empirical evaluation}\label{sec:empirical_evaluation}
In our experiments, we strived to authentically represent IoT devices deployed in an enterprise setting, infected by real-world botnets, and executing genuine attacks. 

\textbf{Lab setup.}
To replicate a typical organizational data flow, we collected the traffic data from IoT devices that were connected via Wi-Fi to several access points, wire connected to a central switch which also connects to a router. 
For sniffing the network traffic, we performed port mirroring on the switch, and recorded the data using Wireshark. 
To evaluate our detection method as realistically as possible, we also deployed all of the components of two botnets (see Figure~\ref{fig:lab_architecture}) in our isolated lab and used them to infect nine commercial IoT devices (see Table~\ref{tab:device_overview_and_params}).

\begin{table*}[!hb]
  \caption{
Overview of the training stage: dataset properties and training summary, optimized hyperparameters for autoencoders, and botnet infections}
  \centering

\resizebox{\textwidth }{!}{
    \begin{tabular}{c||llrcc||cccc||cc}
    \hhline{============}
    & \multicolumn{5}{c}{\textbf{Dataset Properties and Training Summary}} & \multicolumn{4}{c}{\textbf{Optimized Hyperparameters of Autoencoders}} & \multicolumn{2}{c}{\textbf{Botnet Infections}}\\
    \hline
     \begin{tabular}[t]{@{}c@{}}\textbf{Device}\\\textbf{ID}\end{tabular} 
& \textbf{Device Make and Model}         
& \textbf{Device Type}            
& \begin{tabular}[t]{@{}c@{}}\textbf{Number}\\\textbf{of Benign}\\\textbf{Instances}\end{tabular} 
& \begin{tabular}[t]{@{}c@{}}\textbf{Training}\\\textbf{Time}\\\textbf{(\(sec\))}\end{tabular}
& \begin{tabular}[t]{@{}c@{}}\textbf{Object}\\\textbf{Size}\\\textbf{(\(kB\))}\end{tabular}
& \begin{tabular}[t]{@{}c@{}}\textbf{Learning}\\\textbf{Rate}\\\textbf{(\(\eta\))}\end{tabular} 
& \begin{tabular}[t]{@{}c@{}}\textbf{Number}\\\textbf{of Epochs}\\\textbf{(\(epochs\))}\end{tabular} 
& \begin{tabular}[t]{@{}c@{}}\textbf{Anomaly}\\\textbf{Threshold}\\\textbf{(\(tr^*\))}\end{tabular} 
& \begin{tabular}[t]{@{}c@{}}\textbf{Window}\\\textbf{Size}\\\textbf{(\(ws^*\))}\end{tabular}
& \textbf{Mirai}
& \textbf{BASHLITE} \\
    \hhline{============}
    1   & Danmini                & Doorbell        & 49,548   & 555 & 172 & 0.012         & 800 & 0.042     & 82 & \Checkmark               & \Checkmark \\
    2   & Ennio                  & Doorbell        & 39,100   & 215 & 172 & 0.003         & 350 & 0.011     & 22 & - & \Checkmark  \\
    3   & Ecobee                 & Thermostat      & 13,113   & 54 & 172 & 0.028         & 250 & 0.011     & 20 & \Checkmark               & \Checkmark \\
    4   & Philips B120N/10     & Baby Monitor    & 175,240    & 292 & 172 & 0.016         & 100 & 0.030     & 65 & \Checkmark               & \Checkmark \\
    5   & Provision PT-737E        & Security Camera & 62,154  & 275 & 172 & 0.026         & 300 & 0.035     & 32 & \Checkmark          & \Checkmark \\
    6   & Provision PT-838          & Security Camera & 98,514 & 795 & 172 & 0.008         & 450 & 0.038     & 43 & \Checkmark  & \Checkmark \\
    7   & SimpleHome XCS7-1002-WHT & Security Camera & 46,585  & 220 & 172 & 0.017         & 230 & 0.056     & 23 & \Checkmark & \Checkmark  \\
    8   & SimpleHome XCS7-1003-WHT & Security Camera & 19,528 & 190 & 172 & 0.006         & 500 & 0.004     & 25 & \Checkmark & \Checkmark \\
    9   & Samsung SNH 1011 N	            & Webcam          & 52,150 & 150 & 172 & 0.013         & 150 & 0.074     & 32 & - & \Checkmark \\
    \hhline{============}
    \end{tabular}
    }

  \label{tab:device_overview_and_params}
\end{table*}

\textbf{Botnets deployed.}
We focused on two of the most common IoT botnet families: BASHLITE and Mirai. We deployed both of them in our labs and collected traffic data before and after the infection. 

\emph{BASHLITE} (also known as Gafgyt, Q-Bot, Torlus, LizardStresser, and Lizkebab) is one of the most infamous types of IoT botnets, and its code and behavior can be found in other IoT malware as well. 
To launch an attack, the botnet infects Linux-based IoT devices by brute forcing default credentials of devices with open Telnet ports.
In our research, the IoT devices were infected using the binaries from the IoTPOT dataset~\cite{Pa2016IoTPOT:Threats} (namely Gafgyt). 
In order to adjust the attacks to our lab, the IP address of the C\&C server was extracted from the malware's binary, and all of the network traffic to this IP was routed to a server in our lab that functions as a C\&C server. Once a new bot connected to this server and was under its control, this server was able to command the infected device to launch 
attacks.

\emph{Mirai} is the second botent we deployed in our isolated network, using its published source code~\cite{GitHubPurposes}. The experimental setup included a C\&C server and a server with a scanner and loader. 
The scanner and loader components are responsible for scanning and identifying vulnerable IoT devices, and loading the malware to the vulnerable IoT devices detected. 
Once a device was infected, it automatically started scanning the network for new victims while waiting for instructions from the C\&C server.


\begin{figure*}
        \centering
        \begin{subfigure}[b]{0.475\textwidth}
            \centering
            \includegraphics[width=\textwidth]
            {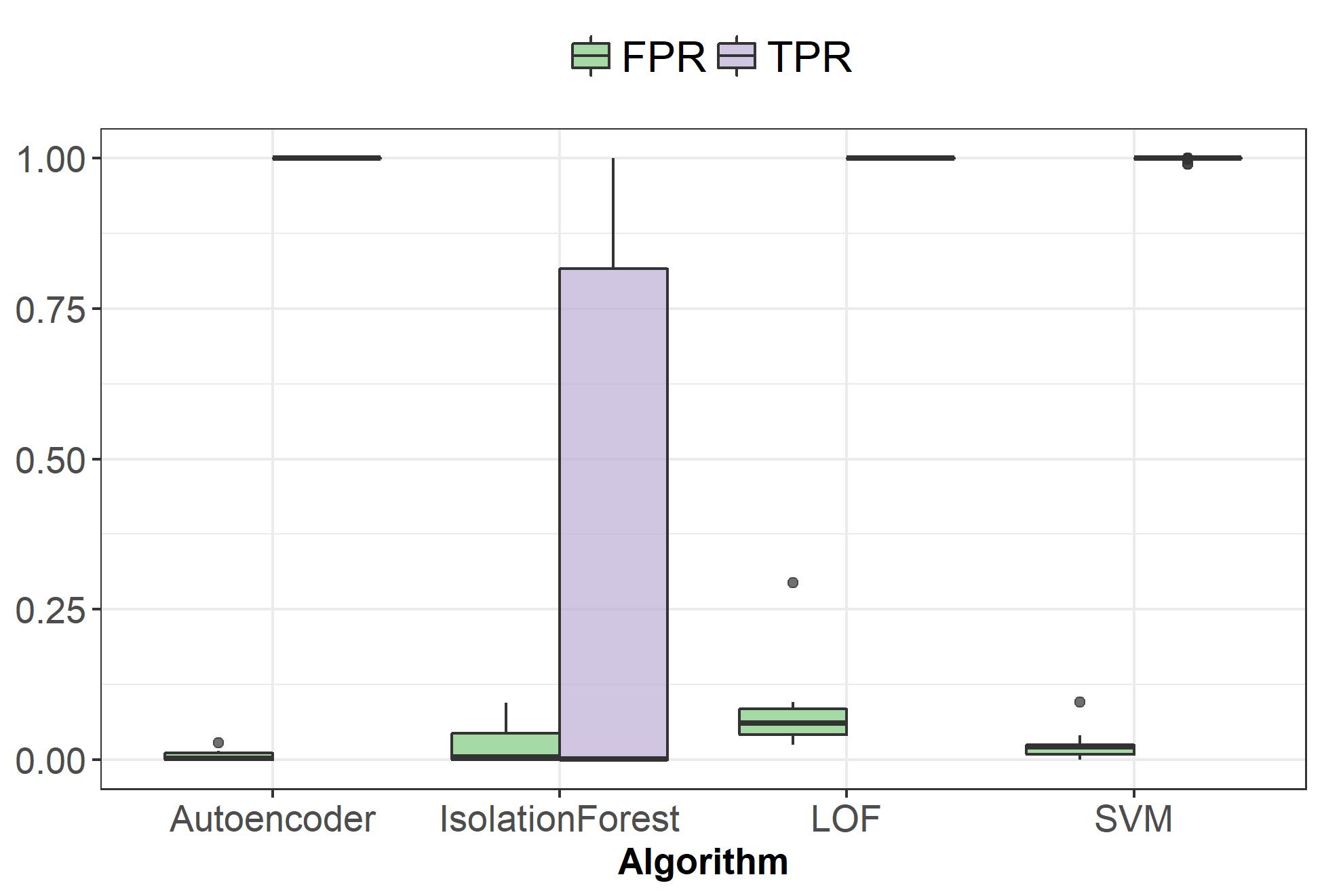}           
            \caption[TPR_FPR_algos]%
            {{\small Methods' detection accuracy}}    
            \label{fig:TPR_FPR_algos}
        \end{subfigure}
        \hfill
        \begin{subfigure}[b]{0.475\textwidth}  
            \centering 
            \includegraphics[width=\textwidth]{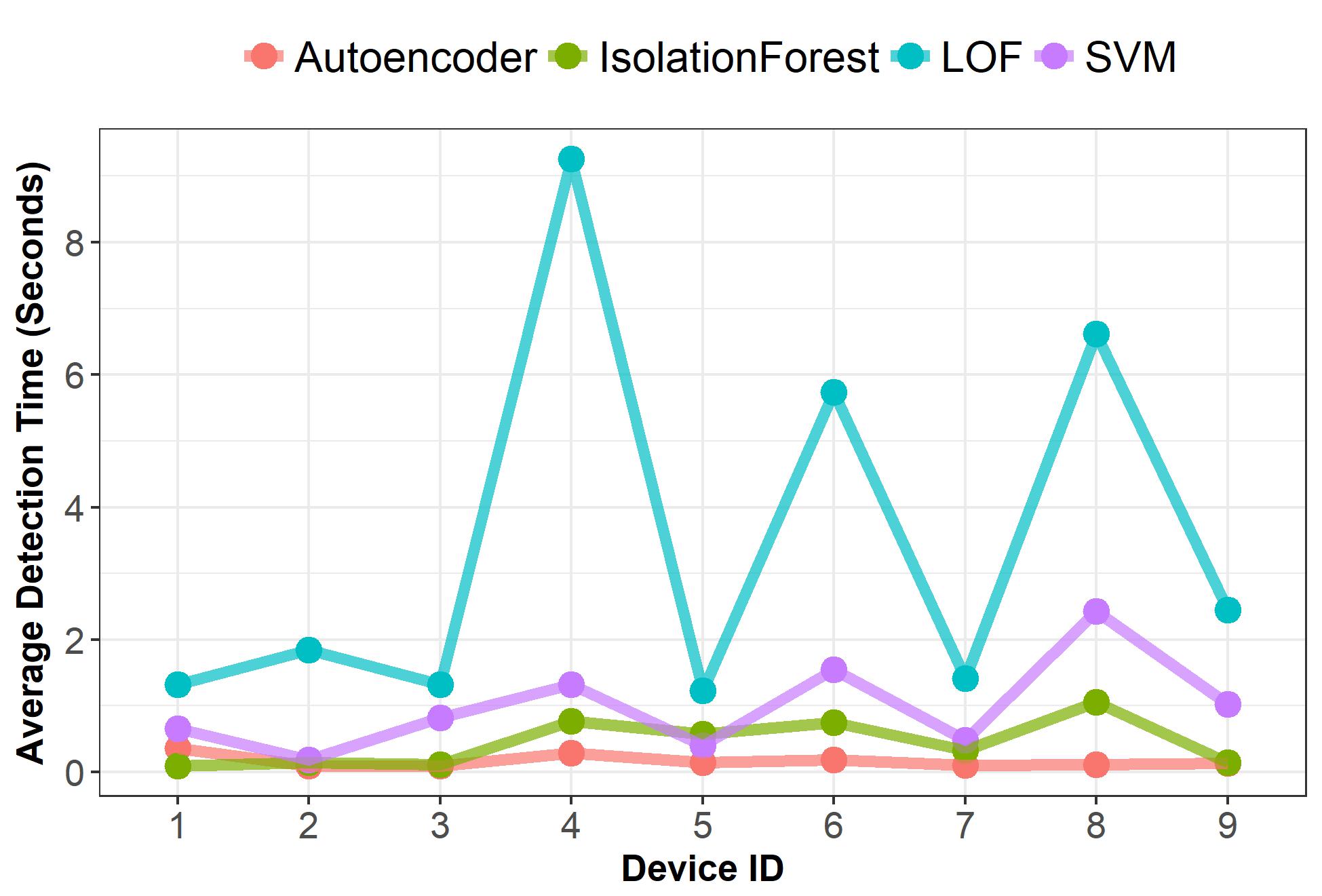}
            \caption[]%
            {{\small Methods' detection time (seconds)}}    
            \label{fig:detection_times_algorithms}
        \end{subfigure}
        \vskip\baselineskip
        \begin{subfigure}[b]{0.475\textwidth}   
            \centering 
            \includegraphics[width=\textwidth
            ]{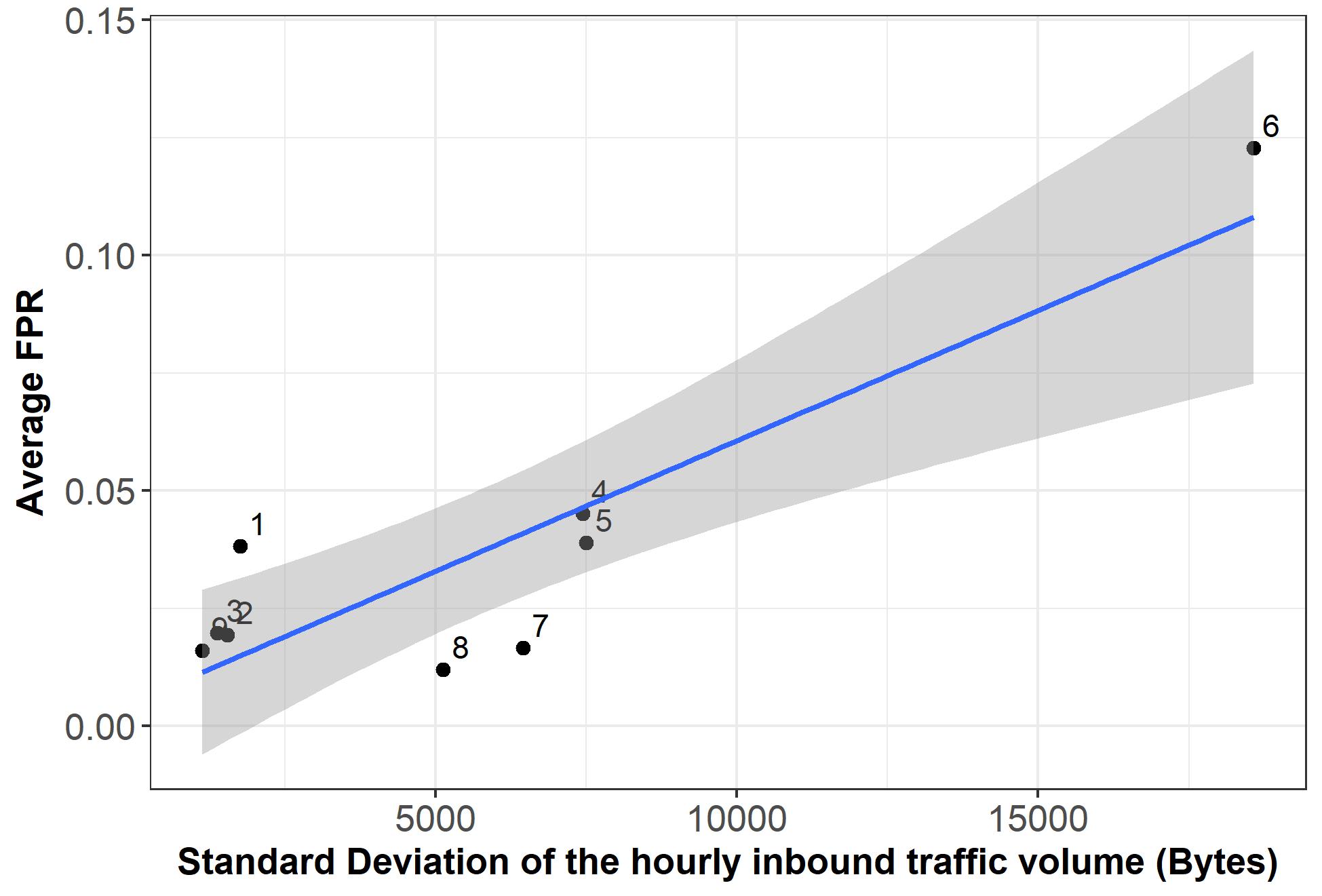}
            \caption[]%
            {{\small Average FPR explained by traffic characteristics}}    
            \label{fig:fpr_d_score}
        \end{subfigure}
        \quad
        \begin{subfigure}[b]{0.475\textwidth}   
            \centering 
            \includegraphics[width=\textwidth
            ]{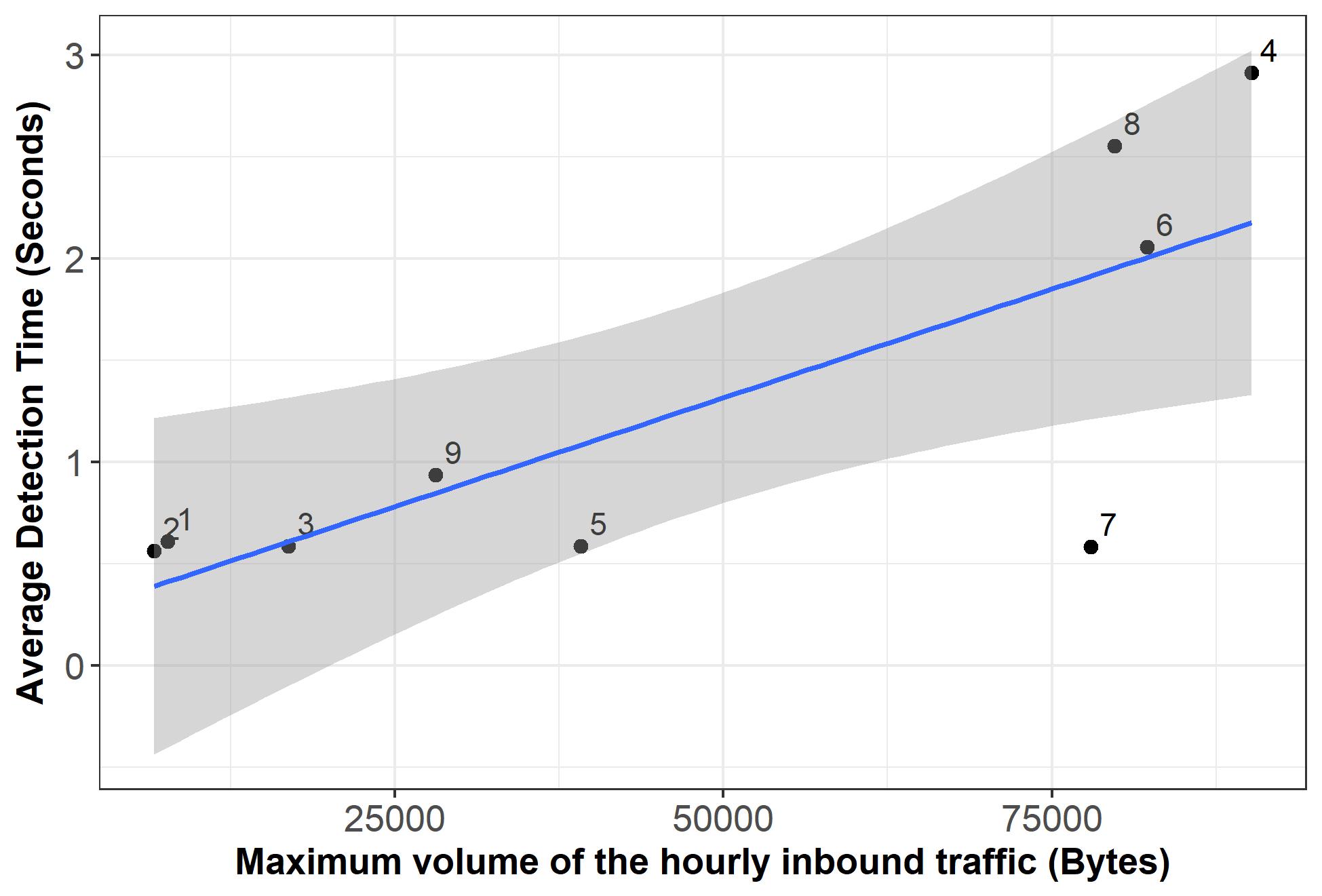}
            \caption[]%
            {{\small Detection time explained by traffic characteristics}}    
            \label{fig:detection_time_d_score}
        \end{subfigure}
        \caption[what is this good for]
        {\small Experimental results using the test set: comparison of methods and potential explanations} 
        \label{fig:results}
    \end{figure*}

\textbf{Attacks executed.} The following is the list of attacks executed and tested in our lab.
\begin{itemize}[leftmargin=*]
	\item BASHLITE Attacks
    \begin{enumerate}
      \item Scan: Scanning the network for vulnerable devices
      \item Junk: Sending spam data
      \item UDP: UDP flooding
      \item TCP: TCP flooding
      \item COMBO: Sending spam data and opening a connection to a specified IP address and port
    \end{enumerate}
    \item Mirai Attacks
    \begin{enumerate}
      \item Scan: Automatic scanning for vulnerable devices
      \item Ack: Ack flooding
      \item Syn: Syn flooding   
      \item UDP: UDP flooding
      \item UDPplain: UDP flooding with fewer options, optimized for higher PPS
    \end{enumerate}
\end{itemize}

\textbf{Experimental results and discussion.}
Each of the nine sets of \emph{benign} data we collected in our lab, corresponding to the nine IoT devices, was divided chronologically into three equidimensional sets: (1) $DS_{trn}$ for training the autoencoder, (2) $DS_{opt}$ for parameter optimization, and (3) the benign part of $DS_{tst}$ for estimating FPR. 
In order to imitate real-world settings and thus assess our method more realistically, we made sure to incorporate traffic from the entire (normal) life cycle of the devices. Particularly, in each of the three sets of each IoT device we included not only traffic data of frequent actions (e.g., a webcam transmitting video) but also infrequent actions (e.g., accessing a webcam via the mobile app, moving in front of it, or booting it).

For training and optimization we used Keras. Each autoencoder had an input layer whose dimension is equal to the number of features in the dataset (i.e., 115). As noted by~\cite{li2015hybrid} and~\cite{arnaldo2017learning}, autoencoders effectively perform dimensionality reduction internally, such that the code layer between the encoder(s) and decoder(s) efficiently compresses the input layer and reflects its essential characteristics. 
In our experiments, four hidden layers of encoders were set at decreasing sizes of 75\%, 50\%, 33\%, and 25\% of the input layer's dimension. 
The next layers were decoders, with the same sizes as the encoders, however with an increasing order (starting from 33\%). 
Table~\ref{tab:device_overview_and_params} provides technical details about the training stage, while focusing on the dataset properties, the optimized hyperparameters of the autoencoders, and the botnet infections.

Following the stage of autoencoder training and optimization, we used the same (benign) data to train three other algorithms commonly used~\cite{tuor2017deep} for anomaly detection: \emph{Local Outlier Factor (\emph{LOF}), \emph{One-Class SVM}, and \emph{Isolation Forest}. 
We optimized their hyperparameters exactly as we did for the autoencoders, including} \(tr\) and \(ws\). 
Finally, we executed all of the above attacks with the same duration via Mirai and BASHLITE's C\&C servers. 
Then we extracted the features from the malicious data and appended each benign part of $DS_{tst}$ (previously mentioned) to the respective malicious part of $DS_{tst}$, to form a single test dataset per IoT device with both benign and malicious instances.  
The experimental results on \(DS_{tst}\) 
(see Figure \ref{fig:results}) are promising:
\begin{itemize}[leftmargin=*]
    \item Our method succeeded in detecting every single attack launched by every compromised IoT device, i.e., TPR of 100\%. As evident in Figure \ref{fig:TPR_FPR_algos}, LOF and SVM reached similar TPRs, much better than the Isolation Forest which demonstrated an inferior and highly variable TPR.
    \item Our method also raised the fewest false alarms. It demonstrated a mean FPR of 0.007$\pm$0.01, lower and more consistent than SVM (0.026$\pm$0.029), Isolation Forest (0.027$\pm$0.041) and LOF (0.086$\pm$0.081).
    \item Moreover, our method required only 174$\pm$212 milliseconds to detect the attacks, and frequently much less time. As evident in Figure~\ref{fig:detection_times_algorithms}, for most of the evaluated IoT devices the average detection time of our method was lower than all the other methods. Assuming that the detection of attack-related anomalies can automatically trigger an immediate isolation of the compromised IoT device from the network, launched attacks can be stopped in less than a second. 
    This is a substantial reduction from the typical duration of DDoS attacks~\cite{Blenn2017QuantifyingBackscatter}, whose distribution normally ranges between 20-90 seconds, plus a long tail where 10\% of the attacks continue more than a day, and 2\% last longer than a month.

\end{itemize} 

In terms of TPR, FPR and detection time, the deep autoencoders exemplified superiority for most devices. 
This is probably due to the ability of deep architectures to learn nonlinear structure mapping and approximate complex functions~\cite{li2015hybrid}. Additionally, the constrained complexity of deep autoencoders, imposed by the reduced dimensionality in the hidden layers, prevents them from learning the trivial identity function~\cite{tuor2017deep}. 
Therefore, deep autoencoders tend to fit common patterns better than uncommon ones. 
This is beneficial for IoT devices, as they normally are task-oriented, so their specified functionality should translate into few normal traffic patterns. 
Despite this tendency to fit common traffic patterns (generated by frequent actions), the autoencoders succeeded in capturing patterns of the infrequent actions (e.g., boots) as well, demonstrated through low FPR. 
In real-world applications, the FPR can be adjusted by manipulating the $tr^*$ and/or $ws^*$, however with some cost of TPR and detection times.

\section{Conclusion}\label{sec:conclusion}
Although the autoencoders in our experiments obtained an FPR of zero on most IoT devices in a test set, the difference in the FPR among the remaining IoT devices led us to further analyze our data. 
We observed that the Philips B120N/10 baby monitor demonstrated the highest FPR relative to the other devices; 
it also produced the largest amount of traffic (see Table~\ref{tab:device_overview_and_params}), so one could expect that the abundance of training instances would result in more robust machine learning models. 
However, this device also has the most diverse set of capabilities, as it is equipped with a two-way intercom function, motion detection, audio detection, and several other sensors for ambient light, temperature, and humidity. 
Given this, it might be more difficult to capture its normal behavior, and therefore future observations may be subject to more categorization errors.

Accordingly, we hypothesize that the difficulty in capturing the normal traffic behavior varies among IoT devices, and that this difficulty may be correlated with (1) the device's capabilities, and (2) the network communications it normally produces. 
A similar notion was raised by~\cite{Bertino2017BotnetsSecurity}, who argues that the specialized functionality of today's IoT devices leads to predictable behaviors. 
In turn, the ease of establishing baseline behaviors for IoT devices facilitates anomaly detection as a means of detecting attacks. To this end, interesting questions arise: 
\begin{itemize}[leftmargin=*]
	\item Can the predictability of traffic behavior of IoT devices be quantified?
    \item Can the relation between the predictability level and the static features of IoT devices (e.g., number and type of sensors, memory size, operating system) or dynamic features (e.g., number of unique destination IPs per hour, variance of the ratio between outgoing and incoming traffic) be formalized?
    \item Can these features be ranked based on their influence on this predictability level?
\end{itemize} 

We presume that the predictability of traffic behavior can be directly translated into performance measures of anomaly detection. 
For example, an IoT device with a high level of traffic predictability would make any anomalous action stand out, and thus the TPR should increase and detection times should decrease in this case. 
For empirical validation we extracted static and dynamic features from the (benign) training set. 
Then we trained regression models to study these features' effect on the average FPR and detection times, obtained on the test set by the four detection methods we evaluated. Figures~\ref{fig:fpr_d_score} and~\ref{fig:detection_time_d_score} depict our preliminary findings via the features found most significant. Figure~\ref{fig:fpr_d_score} shows how an increase in the variability of inbound traffic translates (\emph{p}-value=0.019) into larger average FPR. This makes sense, as lower predictability is prone to manifest through unpredictable (yet benign) traffic behaviors, falsely identified as anomalous. 
Figure~\ref{fig:detection_time_d_score} shows how an increase in the maximal volume of inbound traffic promotes (\emph{p}-value=0.001) longer detection times. 
As we optimize $ws^*$ to reach 0\% FPR on $DS_{opt}$, lower predictability leads to higher $ws^*$ (more instances for majority voting) and subsequently higher detection times.

Ultimately, a solid predictability score can be leveraged by large organizations in order to ensure network functionality and limit the impact that compromised devices might have on the network. 
That is, security policies may not allow the connection of IoT devices with low predictability scores to their networks, since they pose difficulties in attack detection. 
In our future work we plan to further define and investigate the subject of traffic predictability, both theoretically and empirically.

As another extension to the current study, we also plan to evaluate transfer learning techniques by assessing the accuracy of models trained on specific devices when they are applied to identical devices, possibly when connected to other organizational networks. This can help (1) save time (e.g., organizations can deploy models previously learned elsewhere, without the need to collect data and train the models themselves), and (2)	detect compromised IoT devices which have been contaminated prior to connecting to the organizational network, such that the organization has no benign data of them for model training.



\ifCLASSOPTIONcompsoc
  \section*{Acknowledgments}\label{acknowledgments}
\else
  \section*{Acknowledgment}
\fi
The authors would like to 
thank Yan Lin Aung, Amit Subhashchandra Tambe, 
Simon Dzanashvili and Tar Wolfson for their valuable contribution.

\ifCLASSOPTIONcaptionsoff
  \newpage
\fi

\bibliographystyle{IEEEtran}
\bibliography{Mendeley_botnet_no_URL.bib}

\begin{thebibliography}{10}
\providecommand{\url}[1]{#1}
\csname url@samestyle\endcsname
\providecommand{\newblock}{\relax}
\providecommand{\bibinfo}[2]{#2}
\providecommand{\BIBentrySTDinterwordspacing}{\spaceskip=0pt\relax}
\providecommand{\BIBentryALTinterwordstretchfactor}{4}
\providecommand{\BIBentryALTinterwordspacing}{\spaceskip=\fontdimen2\font plus
\BIBentryALTinterwordstretchfactor\fontdimen3\font minus
  \fontdimen4\font\relax}
\providecommand{\BIBforeignlanguage}[2]{{%
\expandafter\ifx\csname l@#1\endcsname\relax
\typeout{** WARNING: IEEEtran.bst: No hyphenation pattern has been}%
\typeout{** loaded for the language `#1'. Using the pattern for}%
\typeout{** the default language instead.}%
\else
\language=\csname l@#1\endcsname
\fi
#2}}
\providecommand{\BIBdecl}{\relax}
\BIBdecl

\bibitem{Kolias2017DDoSBotnets}
C.~Kolias, G.~Kambourakis, A.~Stavrou, and J.~Voas, ``{DDoS in the IoT: Mirai
  and Other Botnets},'' \emph{Computer}, vol.~50, no.~7, pp. 80--84, 2017.

\bibitem{Bertino2017BotnetsSecurity}
E.~Bertino and N.~Islam, ``{Botnets and Internet of Things Security},''
  \emph{Computer}, 2017.

\bibitem{Hallman2017IoDDoSBotnets}
R.~Hallman, J.~Bryan, G.~Palavicini, J.~Divita, and J.~Romero-Mariona,
  ``{IoDDoS The Internet of Distributed Denial of Service Attacks - A Case
  Study of the Mirai Malware and IoT-Based Botnets},'' in \emph{Proceedings of
  the 2nd International Conference on Internet of Things, Big Data and Security
  - Volume 1: IoTBDS}.\hskip 1em plus 0.5em minus 0.4em\relax SciTePress, 9
  2017, pp. 47--58.

\bibitem{Ozcelik2017Software-DefinedDDoS}
M.~Ozcelik, N.~Chalabianloo, and G.~Gur, ``{Software-Defined Edge Defense
  Against IoT-Based DDoS},'' in \emph{2017 IEEE International Conference on
  Computer and Information Technology (CIT)}.\hskip 1em plus 0.5em minus
  0.4em\relax IEEE, 8 2017, pp. 308--313.

\bibitem{Summerville2016Ultra-lightweightDevices}
D.~H. Summerville, K.~M. Zach, and Y.~Chen, ``{Ultra-lightweight deep packet
  anomaly detection for Internet of Things devices},'' in \emph{2015 IEEE 34th
  International Performance Computing and Communications Conference, IPCCC
  2015}, 2016.

\bibitem{Pa2016IoTPOT:Threats}
Y.~M.~P. Pa, S.~Suzuki, K.~Yoshioka, T.~Matsumoto, T.~Kasama, and C.~Rossow,
  ``{IoTPOT: A Novel Honeypot for Revealing Current IoT Threats},''
  \emph{Journal of Information Processing}, vol.~24, no.~3, pp. 522--533, 2016.

\bibitem{Sedjelmaci2016AMethodology}
H.~Sedjelmaci, S.~M. Senouci, and M.~Al-Bahri, ``{A lightweight anomaly
  detection technique for low-resource IoT devices: A game-theoretic
  methodology},'' in \emph{2016 IEEE International Conference on Communications
  (ICC)}.\hskip 1em plus 0.5em minus 0.4em\relax IEEE, 5 2016, pp. 1--6.

\bibitem{Bostani2017HybridApproach}
H.~Bostani and M.~Sheikhan, ``{Hybrid of anomaly-based and specification-based
  IDS for Internet of Things using unsupervised OPF based on MapReduce
  approach},'' \emph{Computer Communications}, 2017.

\bibitem{Butun2015AnomalyThings}
I.~Butun, B.~Kantarci, and M.~Erol-Kantarci, ``{Anomaly detection and privacy
  preservation in cloud-centric Internet of Things},'' in \emph{2015 IEEE
  International Conference on Communication Workshop (ICCW)}.\hskip 1em plus
  0.5em minus 0.4em\relax IEEE, 6 2015, pp. 2610--2615.

\bibitem{Midi2017KalisThings}
D.~Midi, A.~Rullo, A.~Mudgerikar, and E.~Bertino, ``{Kalis — A System for
  Knowledge-Driven Adaptable Intrusion Detection for the Internet of Things},''
  in \emph{2017 IEEE 37th International Conference on Distributed Computing
  Systems (ICDCS)}.\hskip 1em plus 0.5em minus 0.4em\relax IEEE, 6 2017, pp.
  656--666.

\bibitem{Raza2013SVELTE:Things}
S.~Raza, L.~Wallgren, and T.~Voigt, ``{SVELTE: Real-time intrusion detection in
  the Internet of Things},'' \emph{Ad Hoc Networks}, vol.~11, no.~8, 2013.

\bibitem{2017AThings}
B.~B. Zarpelo, R.~S. Miani, C.~T. Kawakani, and S.~C. de~Alvarenga, ``{A survey
  of intrusion detection in Internet of Things},'' \emph{Journal of Network and
  Computer Applications}, vol.~84, pp. 25--37, 4 2017.

\bibitem{tuor2017deep}
A.~Tuor, S.~Kaplan, B.~Hutchinson, N.~Nichols, and S.~Robinson, ``Deep learning
  for unsupervised insider threat detection in structured cybersecurity data
  streams,'' in \emph{Artificial Intelligence for Cybersecurity Workshop at
  AAAI}, 2017.

\bibitem{garcia2014survey}
S.~Garc{\'\i}a, A.~Zunino, and M.~Campo, ``Survey on network-based botnet
  detection methods,'' \emph{Security and Communication Networks}, vol.~7,
  no.~5, pp. 878--903, 2014.

\bibitem{arnaldo2017learning}
I.~Arnaldo, A.~Cuesta-Infante, A.~Arun, M.~Lam, C.~Bassias, and
  K.~Veeramachaneni, ``{Learning Representations for Log Data in
  Cybersecurity},'' in \emph{International Conference on Cyber Security
  Cryptography and Machine Learning}.\hskip 1em plus 0.5em minus 0.4em\relax
  Springer, 2017, pp. 250--268.

\bibitem{li2015hybrid}
Y.~Li, R.~Ma, and R.~Jiao, ``{A hybrid malicious code detection method based on
  deep learning},'' \emph{International Journal of Security and Its
  Applications}, vol.~9, no.~5, 2015.

\bibitem{veeramachaneni2016ai}
K.~Veeramachaneni, I.~Arnaldo, V.~Korrapati, C.~Bassias, and K.~Li, ``Ai\^{} 2:
  training a big data machine to defend,'' in \emph{Big Data Security on Cloud,
  IEEE International Conference on High Performance and Smart Computing (HPSC),
  and IEEE International Conference on Intelligent Data and Security (IDS),
  2016 IEEE 2nd International Conference on}.\hskip 1em plus 0.5em minus
  0.4em\relax IEEE, 2016, pp. 49--54.

\bibitem{yu2017network}
Y.~Yu, J.~Long, and Z.~Cai, ``Network intrusion detection through stacking
  dilated convolutional autoencoders,'' \emph{Security and Communication
  Networks}, vol. 2017, 2017.

\bibitem{GitHubPurposes}
\BIBentryALTinterwordspacing
``{GitHub - jgamblin/Mirai-Source-Code: Leaked Mirai Source Code for
  Research/IoC Development Purposes}.'' [Online]. Available:
  \url{https://github.com/jgamblin/Mirai-Source-Code}
\BIBentrySTDinterwordspacing

\bibitem{Blenn2017QuantifyingBackscatter}
N.~Blenn, V.~Ghi{\"{e}}tte, and C.~Doerr, ``{Quantifying the Spectrum of
  Denial-of-Service Attacks through Internet Backscatter},'' in
  \emph{Proceedings of the 12th International Conference on Availability,
  Reliability and Security - ARES '17}.\hskip 1em plus 0.5em minus 0.4em\relax
  ACM Press, 2017, pp. 1--10.

\end{thebibliography}


\begin{IEEEbiography}[{\includegraphics[width=1in,height=1.25in,clip,keepaspectratio]{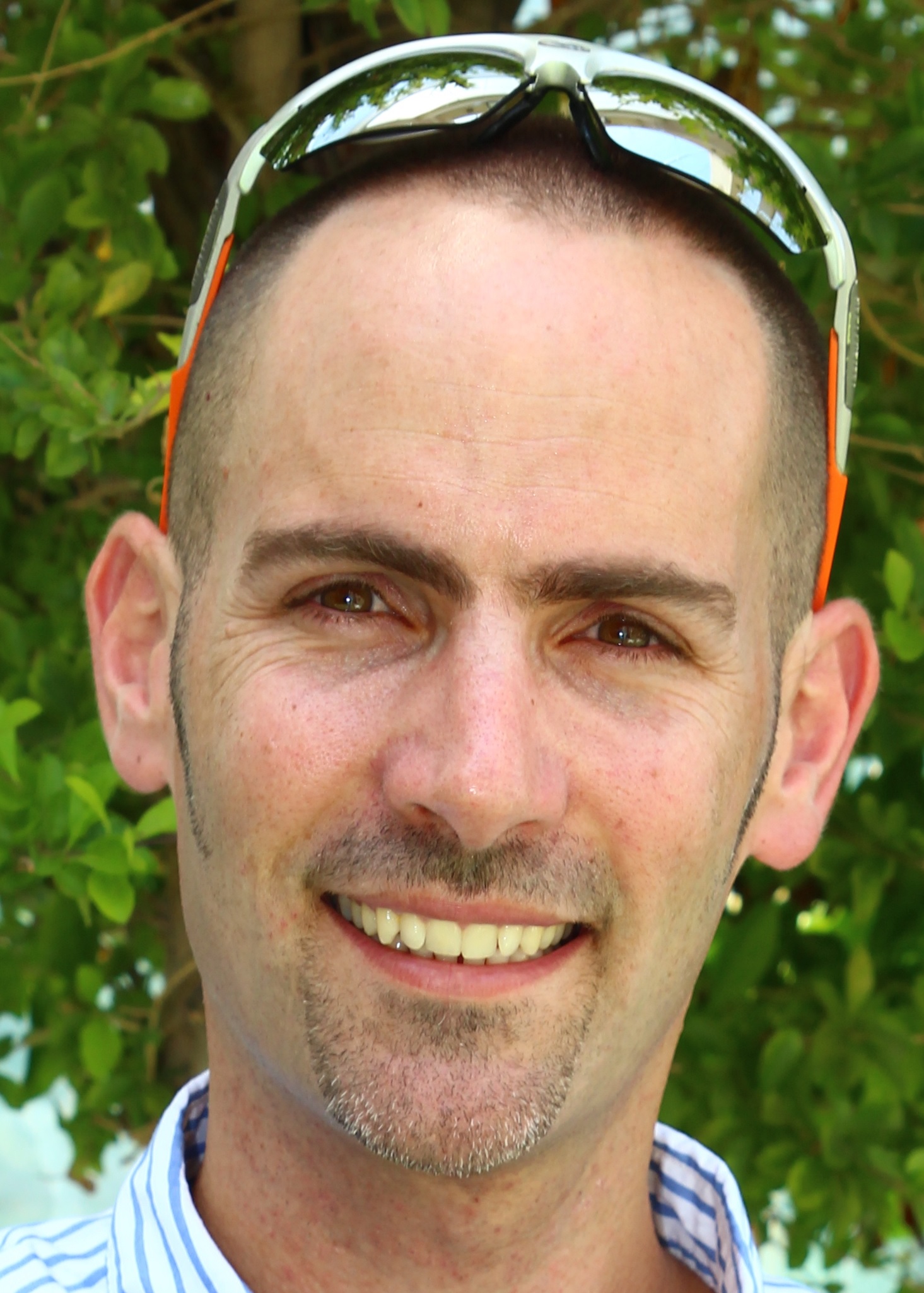}}]{Yair Meidan}is a Ph.D. candidate at the Department of Software and Information Systems Engineering (SISE) at Ben-Gurion University of the Negev (BGU). His research interests include 
machine learning 
and IoT security. Contact him at yairme@post.bgu.ac.il.
\end{IEEEbiography}

\begin{IEEEbiography}[{\includegraphics[width=1in,height=1.25in,clip,keepaspectratio]{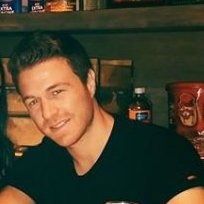}}]{Michael Bohadana}is a M.Sc. student in the SISE Department 
at BGU. 
His research interests include reverse engineering 
and IoT security. Contact him at bohadana@post.bgu.ac.il.
\end{IEEEbiography}

\begin{IEEEbiography}[{\includegraphics[width=1in,height=1.25in,clip,keepaspectratio]{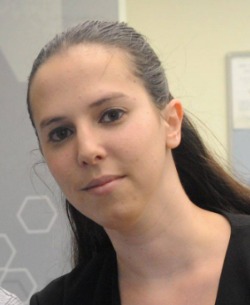}}]{Yael Mathov} is a M.Sc. student in the SISE Department 
at BGU. 
Her research interests include IoT security and reverse engineering.
Contact her at yaelmath@post.bgu.ac.il.
\end{IEEEbiography}

\begin{IEEEbiography}[{\includegraphics[width=1in,height=1.25in,clip,keepaspectratio]{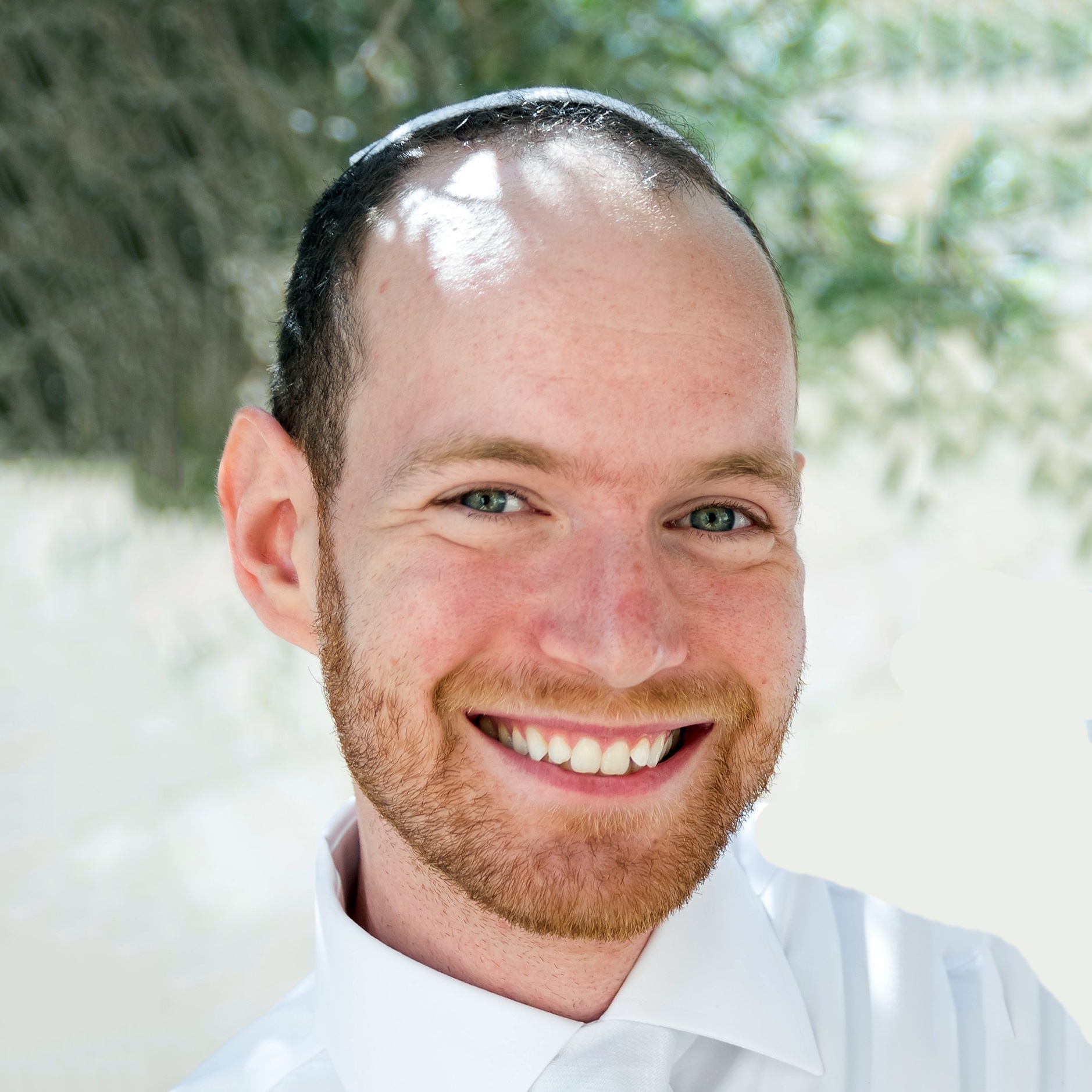}}]{Yisroel Mirsky}is a Ph.D. candidate in the SISE Department 
at BGU. 
His research interests include machine learning and time series anomaly detection. Contact him at yisroel@post.bgu.ac.il.
\end{IEEEbiography}

\begin{IEEEbiography}[{\includegraphics[width=1in,height=1.25in,clip,keepaspectratio]{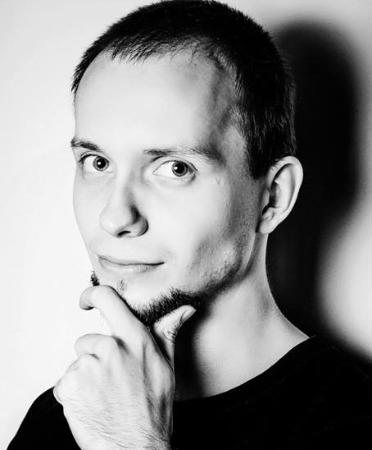}}]{Dominik Breitenbacher} 
is a research assistant at iTrust Centre of Cybersecurity at Singapore University of Technology and Design (SUTD). His research interests include IoT security and malware detection. Contact him at dominik@sutd.edu.sg.
\end{IEEEbiography}

\begin{IEEEbiography}[{\includegraphics[width=1in,height=1.25in,clip,keepaspectratio]{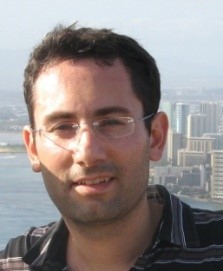}}]{Asaf Shabtai}is an assistant Professor in the SISE Department 
at BGU. 
His research 
interests include 
computer and network security, and machine learning. 
Contact him at shabtaia@bgu.ac.il.
\end{IEEEbiography}




\begin{IEEEbiography}[{\includegraphics[width=1in,height=1.25in,clip,keepaspectratio]{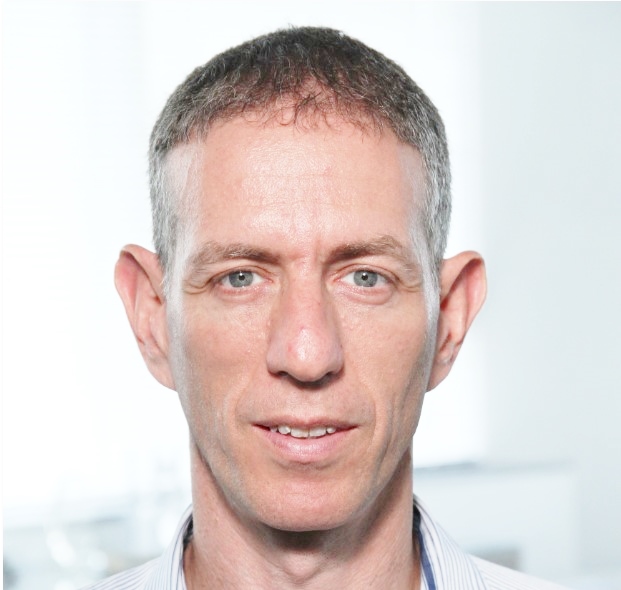}}]{Yuval Elovici}is the director of the Telekom Innovation Laboratories at BGU, 
head of the Cyber Security Research Center at BGU, research director of iTrust at SUTD, 
and a Professor in the SISE Department 
at BGU. His 
research interests include 
computer and network security, 
and machine learning. Contact him at yuval\_elovici@sutd.edu.sg.
\end{IEEEbiography}

\end{document}